\documentclass[graybox]{svmult}

\usepackage{mathptmx}       
\usepackage{helvet}         
\usepackage{courier}        
\usepackage{type1cm}        
  \usepackage{subfigure}
\usepackage{makeidx}         
\usepackage{multicol}        
\usepackage[bottom]{footmisc}
\usepackage{subfigure}
\usepackage{url}

\usepackage[pdftex]{graphicx}

\makeindex             

\begin{document}

\title*{A Risk Based approach for the Solvency Capital requirement for Health Plans}
\author{Fabio Baione Davide Biancalana and Paolo De Angelis}
\institute{Fabio Baione, \at Department of Statistics, Sapienza University of Rome, Viale Regina Elena, 295 00161 Roma, Tel. +39 06 49255317, \email{fabio.baione@uniroma1.it}
\and Davide Biancalana, \at Department of Statistics, Sapienza University of Rome, Italy \email{davide.biancalana@uniroma1.it}
\and Paolo De Angelis, \at Department of methods and models for economics territory and finance, Sapienza University of Rome, Italy \email{paolo.deangelis@uniroma1.it}
}
\maketitle
\abstract{The study deals with the assessment of risk measures for Health Plans in order to assess the Solvency Capital Requirement. For the estimation of the individual health care expenditure for several episode types, we suggest an original approach based on a three-part regression model. We propose three Generalized Linear Models (GLM) to assess claim counts, the allocation of each claim to a specific episode and the severity average expenditures respectively. One of the main practical advantages of our proposal is the reduction of the regression models compared to a traditional approach, where several two-part models for each episode types are requested. As most health plans require co-payments or co-insurance, considering at this stage the non-linearity condition of the reimbursement function, we adopt a Montecarlo simulation to assess the health plan costs. The simulation approach provides the probability distribution of the Net Asset Value of the Health Plan and the estimate of several risk measures.}

\section{Introduction}
\label{sec:1}
The Italian National Health System (SSN) is based on three pillars. In particular, the second is mainly characterized by private group health plans and usually provided through labor agreements. However, the lack of a clearly defined authority and solvency requirements raises insolvency risk, particularly for self-insured funds. Our aim is to introduce an actuarial global framework that allows an estimation of a short or medium term solvency capital requirement.
The first step consists of the prediction of the one-year health care expenditures at an individual level for several episode types, with an original approach, alternative to the ones in~\cite{Duncan} and~\cite{Frees}. Then, considering deductibles, copayments and other limitations working on single episode, or single person or family level, we focus on the estimate of the reimbursement amount. Our final goal is the estimate of the density function of health plan profit (losses) and revenues by a simulation technique to calculate the solvency capital requirement according to several risk measures.

\section{Actuarial Framework}
We consider an Health Plan  (henceforth HP) composed by $r$ policyholders.
Let
\begin{itemize}
	\item $i$ index the $i$-th policyholder, $1\leq i \leq r$;
	\item $j$ index the $j$-th branch of health expenditure, $1\leq j \leq J$;
	\item $h$ index the $h$-th family, $1\leq h \leq H$
	\end{itemize}
In the following we denote with
\begin{itemize}
	\item $N$  the random variable (r.v.), number of episodes per year;
	\item $T$  the r.v. branch of the episode requested;
	\item $Y$  the r.v. expenditure for single episode;
	\item $Z$  the r.v. expenditure for single policyholder per year;

\end{itemize}
In classical individual risk model, the expenditure of the $i$-\emph{th} policyholder given a specific branch $j$ is
\begin{equation}
   \label{Zij}
   Z_{i,j}=\sum_{g=1}^{N_{i,j}}Y_{i,j,g}
\end{equation}
where $Y_{i,j,g}$ is the r.v. expenditure for $i$-th insured, $j$-th branch and $g$-th episode.
The total expenditure of the HP is
\begin{equation}
   \label{Z}
   Z=\sum_{i=1}^{r}\sum_{j=i}^{J}Z_{i,j}\,.
\end{equation}

Under the typical assumption of independence between $N_{i,j}$ and $Y_{i,j,g}$ and identical distribution on $Y_{i,j,g}, \forall g=1,\ldots,N_{i,j}$, then
\begin{equation}
   E\left[Z_{i,j}\right] =E\left[ N_{i,j}\right]\cdot E\left[ Y_{i,j}\right]\,.
\end{equation}

It is worth nothing that
\begin{equation}
   \label{dec}
   E\left[N_{i,j}\right] =E\left[ N_{i}\right]\cdot Prob\left[ T_i=j|N_i>0\right]\,,
\end{equation}	
then the total expenditure for single insured is
\begin{equation}
   E\left[Z_i\right] = \sum_{j=1}^{J} E\left[ Z_{i,j}\right]=\sum_{j=1}^{J} E\left[ N_{i}\right]\cdot Prob\left[ T_i=j|N_i>0\right]\cdot E\left[ Y_{i,j}\right].
\end{equation}	

It is possible to calculate the expenditure for each family and single branch
\begin{equation}
   Z_{h,j} = \sum_{i\in h}^{} Z_{i,j}\,,
\end{equation}
then $E\left[ Z_{h,j}\right]  = \sum_{i\in h}^{} E\left[ Z_{i,j}\right] $ and $E\left[ Z_{h}\right]  = \sum_{j=1}^{J} E\left[ Z_{h,j}\right] $. Hence, the expected total expenditure for the health fund is
\begin{equation}
   \label{E(Z)}
   E\left[Z\right] = \sum_{h=1}^{H} E\left[ Z_h\right]=\sum_{i=1}^{r} E\left[ Z_i\right]\,.
\end{equation}

In order to assess the reimbursement of the HP, we introduce the following notation
\begin{itemize}
	\item $f_j$ is the deductible for single episode for $j$-th branch;
	\item $s_j$ is the coinsurance for single episode for $j$-th branch;
	\item $M_j$ is the out of pocket maximum for single episode for $j$-th branch;
	\item $M_j^\ast$ is the out of pocket maximum for family for $j$-th branch;
	\item $L$ is the r.v. reimbursement for single episode;
	\item $K$ is the r.v. reimbursement for single policyholder per year.	
\end{itemize}


Then, the reimbursement for the single episode $g$ of the $i$-th insured and $j$-th branch is
\begin{equation}
   \label{Lij}
   L_{i,j,g} = min\left\lbrace Y_{i,j,g}-max\left[ s_j\cdot Y_{i,j,g};f_j\right] ;M_j\right\rbrace \,.
\end{equation}

Following (\ref{Zij}), the reimbursement of a single policyholder $i$ for a specific branch $j$ is given by
\begin{equation}
   \label{Kij}
   K_{i,j}=\sum_{g=1}^{N_{i,j}}L_{i,j,g}\,.
\end{equation}

Considering that there is a $M_j^\ast$ for each branch $j$, that works on the single family, it is important to assess the r.v. reimbursement in terms of family as follows
\begin{equation}
   \label{Khj}
   K_{h,j}= min\left( \sum_{i\in h}^{}K_{i,j};M_j^\ast\right)\,.
\end{equation}

Hence, the total reimbursement should be computed only for family aggregation as follows
\begin{equation}
   \label{K}
   K=\sum_{h=1}^{H}\sum_{j=i}^{J}K_{h,j}\, .
\end{equation}

Under the typical assumption of independence between $N_{i,j}$ and $L_{i,j,g}$ and identical distribution on $L_{i,j,g}, \forall g=1\ldots,N_{i,j}$, then
\begin{equation}
   E\left[K_{i,j}\right] =E\left[ N_{i,j}\right]\cdot E\left[ L_{i,j}\right]= E\left[ N_{i}\right]\cdot Prob\left[ T_i=j|N_i>0\right]\cdot E\left[ L_{i,j}\right].
\end{equation}

Whereas, the expected reimbursement for the $h$-\emph{th} family is
\begin{equation}
   E\left[ K_{h,j}\right] =E\left[ min\left( \sum_{i\in h}^{}K_{i,j};M_j^\ast\right) \right]\,.
\end{equation}
Hence, the one year expected total loss for the health fund is
\begin{equation}
   E\left[K\right] = \sum_{h=1}^{H}\sum_{j=1}^{J} E\left[ K_{h,j}\right]\,.
\end{equation}

Once defined the one year loss of the HP it is important to define contribution for the same time period. Assuming that each policyholder has to pay a fixed contribution $b$, the total cash in amount is $C=b\cdot r$. Hence the r.v one year profit is
\begin{equation}
   \label{profit}
    U = C-K
\end{equation}
and
\begin{equation}
   \label{profit}
   E[U] = C-E[K]\,.
\end{equation}

It is worth noting, that the one year total amount of contribution is deterministic.
The actuarial framework represented refers to the expected values of r.v. considered.
Hence, for the estimate of the expected value of r.v. $Z$, is sufficient to choose a statistical method of point estimation for $E\left[N_{i,j}\right]$, $E\left[Y_{i,j}\right]$ and $Prob\left[ T_i=j|N_i>0\right]$  such as likelihood or method of moments. Then, all the expected values of $Z$ for different aggregations are obtained by sum, since the expected value is a linear operator.
It is important to note that equations (\ref{Lij}) and (\ref{Khj}) introduce a non linear relation between the random variables involved due to the effect of limitations.
This implies that to define the expected value of r.v. $K$ and consequently $U$, an estimate of the density function (df) is necessary.\\
Moreover, if we are interested on the assessment of a specific risk measure, we need to consider the relative properties of the assumed risk measure (e.g. sub-additivity, homogeneity, etc.). This implies that also for the calculation of $Z$, and a fortiori for $K$ and $U$, it is fundamental to estimate such a density function.\\
Given the complexity of the stated probability structure, the use of a simulation approach is necessary.

\section{Numerical Investigation}

We set our framework on a database from an Italian HP between years 2009 and 2013. The portfolio has $r=53,984$ policyholders, and $H=24,660$ families. The number of observed episodes is 341,494 spread in 21 branches.
In order to estimate the df of $U$ we need to start by a specific probabilistic structure for the main r.v.s:
\begin{itemize}
	\item $N_{i}$ is Negative Binomial distributed;
	\item $\left[ T_i|N_i>0\right] $ multinomial distributed;
	\item $Y_{i,j}$ is Gamma distributed;
\end{itemize}
The subscripts $i$ and $j$ means that we assume a specific distribution for each policyholder and branch, respectively. To this aim, a possible choice is the introduction of a dependency structure between the response variables and a set of covariates by means of a regression model. Considering the features of the r.v.s previously introduced Generalized Linear Models (see~\cite{Nelder}) (GLM) seem an appropriate choice.
Considering a vector of covariates $\boldmath{x}_i$ for each policyholder, by means of GLM we can estimate the conditional mean of $E\left[N|\boldmath{x}_i\right]=E\left[N_i\right]$,
 $E\left[T_i=j|N_i>0\right]$ and $E\left[Y\left| \boldmath{x}_i,j\right.\right]=E\left[Y_{i,j}\right]$.
In Figure~\ref{Fig: E(Yij)}, we introduce an example of fitting analysis for the three GLM models based on the analysis over age of a Male for the branch "dental visit":

\begin{figure}[!ht]
\sidecaption[t]
	\centering
	\subfigure[Expected number of episodes]{
		\includegraphics[angle=0, width=0.45\textwidth]{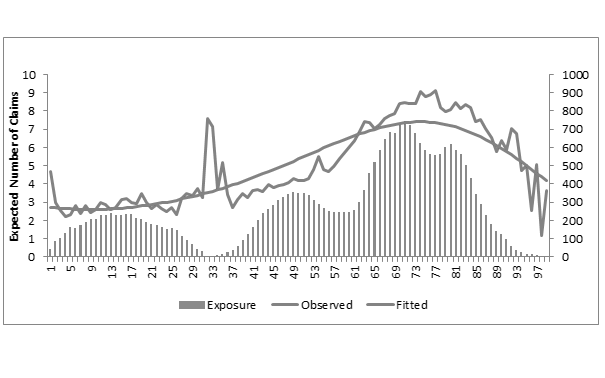}}
	\centering
	\subfigure[Probability of a dental visit]{
		\includegraphics[angle=0, width=0.45\textwidth]{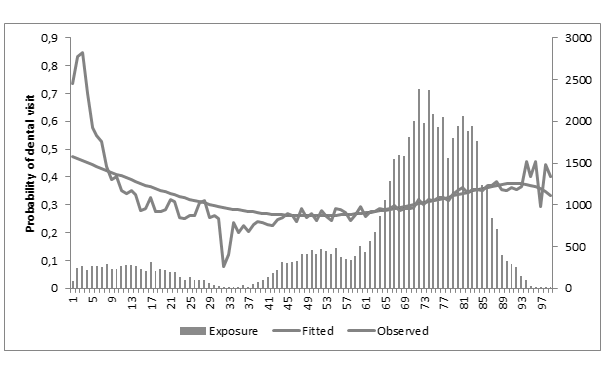}}
	\centering
	\subfigure[Expected expenditure per episode]{
		\includegraphics[angle=0, width=0.45\textwidth]{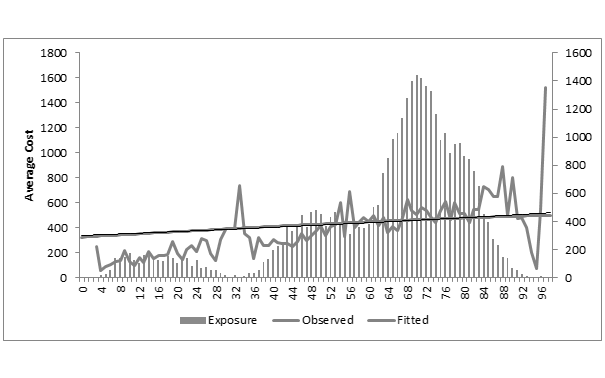}}
	\caption{Goodness of fitting.}
	\label{Fig: E(Yij)}
\end{figure}

As one can observe in Fig.\ref{Fig: E(Yij)}.(a), the expected number of claims increases as the age increases; whereas, in Fig.\ref{Fig: E(Yij)}.(b) for  $Prob\left[T_i=j|N_i>0\right]$, a U-shape is observable as a consequence of the higher request of orthodontics implant in young ages and of the increase due to the dental disease at older ages. The expenditure for single episode has only a small increasing trend, however age is not a significant rating factor as observable in Fig.\ref{Fig: E(Yij)}.(c).
As previously stated, the assessment of the expected values of the r.d.s $N, T|N$ and $Y$ is only determinant for the calculation of the expected total expenditure $E(Z)$ (see Eq.~\ref{E(Z)}), therefore for the estimation of the expected total amount of reimbursement $E(K)$ a simulation approach is necessary.\\
By exploiting the assumed distributions and the capability of the GLM to provide the full set of parameters to specify the conditional density functions, we carry out a Monte Carlo simulation whose final result is the estimate of all involved r.v.s density functions and especially the profit r.v. $U$.\\
The final outcome is a sampled distribution on which is possible to assess the Solvency capital requirement according to several risk measures and/or a specific regulatory framework.

In Fig~\ref{df} an example of the sample distribution of the r.v. $U$ is reported:
\begin{figure}[!htbp]
	\sidecaption[t]
	\centering
	\includegraphics [scale=0.38] {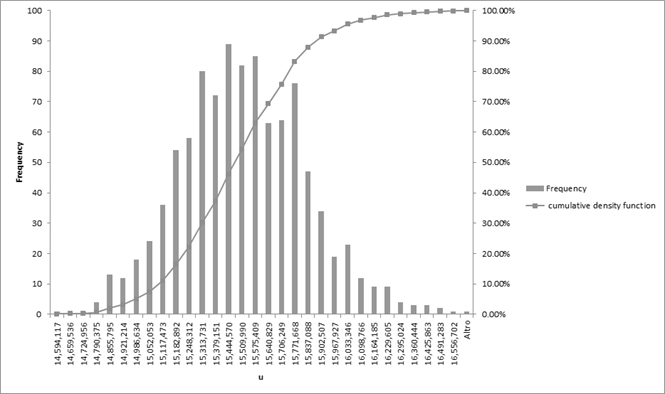}
	\caption{Density function of r.v. $U$.} \label{df}
\end{figure}


\begin{thebibliography}{99.}%
%
%

%
\bibitem{Duncan}  Duncan, I. Loginov, M., Ludkovski,M.:  Testing Alternative Regression Frameworks for Predictive Modeling of Health Care Costs. North American Actuarial Journal \textbf{20(1)}, 65-87 (2016)
\bibitem{Frees} Frees, E.W., Lee,G.Y., Rosenberg, M.A. : Predicting the Frequency and Amount of Health Care Expenditures. North American Actuarial Journal \textbf{15(3)}, 377-392 (2011)
\bibitem{Nelder} Mc Cullagh, P.A., Nelder, J.A.: Generalized Linear Models. Taylor \& Francis, USA (1989)



\end{thebibliography}
\end{document}